\documentstyle[11pt]{article}

\begin{document}

\title{The Custodial Symmetry}
\author{Rodolfo A. Diaz and R. Mart\'{\i}nez \\
Depto. de F\'{\i}sica\\
Universidad Nacional \\
Bogot\'a, Colombia}
\date{}
\maketitle

\begin{abstract}
In the present work, we elucidate the meaning of the custodial symmetry and
its importance at the phenomenological level in the framework of the
standard model of the electroweak interactions and its possible extensions.
\end{abstract}

\section{Introduction}

The Standard Model (SM) of the strong and electroweak interactions which has
the $SU(3)_{C}\otimes SU(2)_{L}\otimes U(1)_{Y}$ gauge group symmetry\cite
{uno} has been quite succesful. Its predictions are in excellent agreement
with the experimental data. Moreover, the model has predicted new particles
like the $Z$ gauge boson which is the quantum of the weak interaction with
zero electric charge. It has also predicted the charm quark which was
introduced to avoid the flavor changing neutral currents at tree level\cite
{dos}, and top quark. However, the SM has some problems. It does not unify
the strong and electroweak interactions and has three coupling constants
associated to each gauge group. Currently there are some theories which
unify these interactions and they only contain one coupling constant and a
simple Lie group; however, they predict a proton decay faster than the
experimental limit. Some of these groups are: $SU(5)$, $SU(6)$, $SO(10)$, $%
E_{6}$ etc\cite{tres}.

One of the most interesting problems of the SM is that there is no
explanation of the origin of masses, mixing of the fermions and the number
of families appears as an identical pattern, although there are some models
which introduce horizontal symmetries to explain the relation of the fermion
masses by spontaneous symmetry breaking (SSB). The physicists expect to find
a symmetry which relates the fermion generations; the general aim is to
distinguish among them by breaking this symmetry and generating hierarchical
mass scales that give different radiative corrections to the fermions. This
symmetry would commute with the SM or a simple group that contains it. Some
of the simple gauge groups that unify the SM with horizontal symmetries are: 
$SU(11)$, $SO(14)$, $SO(18)$, $E_{8}$, etc\cite{cuatro}. At the present time
there is not a realistic model that can {\it explain satisfactorily} the
problem of the fermion masses.

To give mass to the fermions and the gauge bosons in the SM, it is necessary
to introduce a multiplet of scalar fields which has some components that are
`eaten` by the gauge bosons, the would-be Goldstone bosons, and another
component which is called the Higgs particle. This is a physical degree of
freedom, but the theory cannot predict its mass. There are some energy
regions to look for the Higgs according to the production and decay
channels. Up to now there is not evidence of it.

The left (right) quirality fermions in the SM transform according to the
global symmetry $SU(2)_{L(R)}$ and the Higgs doublet, four real fields, is a
bidoublet under this global symmetry. Before the breaking of $%
SU(2)_{L}\otimes U(1)_{Y}$ , the Higgs potential has a global $%
SU(2)_{L}\otimes SU(2)_{R}$ symmetry which reduces to $SU(2)_{V}\;$when the
symmetry is broken. This remanent global symmetry is called the ''custodial
symmetry''\cite{cinco}. Additionally, it is possible to write down the
Yukawa Lagrangian to see the same $SU(2)_{L}\times SU(2)_{R}$ global
symmetry explicitly.

In the present work, we explain the meaning of the custodial symmetry
specially when dealing with Higgs multiplets appearing in extensions of the
SM. We also explain that if the gauge symmetry of the electroweak model is
broken by the Higgs doublet, there is a custodial symmetry which protects
the mass relation of the $W$ and $Z$ gauge fields, i. e., 
\begin{equation}
\rho \equiv \frac{M_{W}^{2}}{M_{Z}^{2}\cos ^{2}\theta }=1.
\end{equation}
This relation is also valid after the radiative corrections if the custodial
symmetry is not broken. Experimentally this relation is satisfied at the $%
1\% $ level, which restricts the new physics beyond the SM and allows us to
distinguish among different models.

\section{The gauge field masses}

The kinetic energy parts of the Lagrangian of the scalar fields produce the
mass of the gauge fields after the SSB. If we have an arbitrary scalar
multiplet of $SU(2)_{L}\otimes U(1)_{Y}$, the gauge field masses arise from 
\[
D_{\mu }<\phi >_{o}\longrightarrow ig\vec{A_{\mu }}\cdot \lbrack \;\vec{T}%
,<\phi >_{o}]+i\frac{g^{\prime }}{2}\;B_{\mu }\;[\;Y,<\phi >_{o}] 
\]
where $<\phi >_{o}$ is the vacuum expectation value (v.e.v.) of the scalar
multiplet, i.e., $<\!\phi \!>_{o}=<\!0|\phi |0\!>$. The covariant derivative
for a general representation of $SU(2)_{L}\otimes U(1)_{Y}$ can be written
as 
\begin{eqnarray}
D_{\mu }<\phi >_{o}\longrightarrow \!\! &i&\!\!\frac{g}{\sqrt{2}}\;W_{\mu
}^{+}\;[\;T^{+},<\phi >_{o}]+i\frac{g}{\sqrt{2}}\;W_{\mu
}^{-}\;[\;T^{-},<\phi >_{o}]  \nonumber \\
+\!\! &ig&\!A_{3\mu }\;[\;T_{3L},<\phi >_{o}]+i\frac{g^{\prime }}{2}\;B_{\mu
}\;[\;Y,<\phi >_{o}].
\end{eqnarray}
where we have defined $W_{\mu }^{\pm }=(A_{\mu }^{1}\mp iA_{\mu }^{2})/\sqrt{%
2}$.

The SM has a $U(1)_{Q}$ remanent symmetry and $<\phi >_{o}$ is different
from zero because of the neutral component, so that we get 
\begin{equation}
T_{3L}<\phi >_{o}=-\frac{1}{2}Y<\phi >_{o}.
\end{equation}
Using the Lie algebra, the mass terms of the gauge fields can be written as 
\cite{cheng} 
\begin{equation}
\left( \frac{g^{2}}{2}\left[ t(t+1)-t_{3L}^{2}\right] \left( A_{\mu
}^{1}A^{1\mu }+A_{\mu }^{2}A^{2\mu }\right) +t_{3L}^{2}\left( gA_{3\mu
}-g^{\prime }B_{\mu }\right) ^{2}\right) <\phi >_{o}^{2}\;,
\end{equation}
where $t(t+1)$ and $t_{3L}$ are the eigenvalues of the $%
\sum_{i=1}^{3}T_{iL}^{2}$ and $T_{3L}$ operators, respectively. Under the
assumption that $g^{\prime }=0$ or equivalently $\sin \theta _{W}=0,$ and
that the $SU(2)_{L}$ gauge fields transform as a triplet of the $SU(2)$
global symmetry, and the kinetic energy Lagrangian is invariant under these
transformation, the following relation is satisfied 
\begin{equation}
t(t+1)-t_{3L}^{2}=2t_{3L}^{2}.  \label{eqww}
\end{equation}
This symmetry is only exact when the $B_{\mu }$ fields is zero because it
breaks the custodial symmetry.

From the above relations and the definition of the mass terms 
\begin{eqnarray}
M_{W}^{2}\! &=&\!g^{2}\left( t(t+1)-t_{3L}^{2}\right) <\phi >_{o}^{2}\;, \\
M_{Z}^{2}\! &=&\!2\;(g^{2}+g^{^{\prime }2})\;t_{3L}^{2}<\phi >_{o}^{2}, 
\nonumber
\end{eqnarray}
for the $\rho $ parameter we get 
\begin{eqnarray}
\rho \! &=&\!\frac{M_{W}^{2}}{M_{Z}^{2}\cos ^{2}\theta _{W}}  \nonumber \\
&=&\frac{t(t+1)-t_{3L}^{2}}{2t_{3L}^{2}}\frac{g^{2}}{\left( g^{2}+g^{\prime
}{}^{2}\right) }\frac{1}{\cos ^{2}\theta _{W}}  \nonumber \\
&=&1
\end{eqnarray}
where $\tan \theta _{W}=g^{^{\prime }}/g$ and $Z_{\mu }=\cos \theta
_{W}A_{3\mu }-\sin \theta _{W}B_{\mu }$. Taking into account the eq.(\ref
{eqww}), we can get different values for $t$ when $\rho =1$ 
\begin{eqnarray}
t &=&\frac{1}{2}\left( -1+\sqrt{1+12t_{3L}^{2}}\right)  \nonumber \\
&=&0,\;\frac{1}{2},\;3,...\;\;.
\end{eqnarray}
The first and second solutions correspond to singlets and doublets,
respectively. The other solutions are not important in the SM because they
do not couple to the fermions in the Yukawa Lagrangian and they do not
produce fermion mass terms. However, for other fermion representations,
physics beyond the SM, Higgs fields with other values of $t$ are important
to construct Yukawa terms. One interesting example is the see-saw mechanism
implemented with new fermions and triplet Higgs fields. In the minimal SM
with only left handed neutrino, it is neccesary a Higgs triplet to get a
massive neutrino; however the vev of the new Higgs representation is
restricted by the experimental value of the $\rho $ parameter. With the
above arguments we can say that if we use doublets to break the SM, the $%
SU(2)_{L}$ gauge fields will be a triplet of the $SU(2)$ global symmetry or
the custodial symmetry and $\rho $ is equal to one. Radiative corrections to
the $\rho $ parameter by gauge fields violate softly this symmetry.

\section{The Higgs and fermions Lagrangian}

The most general Higgs potential which is renormalizable and invariant under 
$SU(2)_{L}\otimes U(1)_{Y}$ gauge transformations have the form 
\begin{equation}
V=\lambda \left( \phi ^{\dagger }\phi -\mu ^{2}\right) ^{2}\;,
\end{equation}
where 
\[
\phi =\frac{1}{\sqrt{2}}\left( 
\begin{array}{l}
\phi _{1}+i\phi _{2} \\ 
\phi _{4}+i\phi _{3}
\end{array}
\right) . 
\]
The potential as function of the $\phi _{i}$ scalar fields can be written as 
\begin{equation}
V=\frac{\lambda }{4}\left( \phi _{1}^{2}+\phi _{2}^{2}+\phi _{3}^{2}+\phi
_{4}^{2}-2\mu ^{2}\right) ^{2}.
\end{equation}
It is invariant under rotations of the four fields which lead to $SO(4)$ as
the global symmetry group. This group is isomorphic to $SU(2)_{L}\otimes
SU(2)_{R}$ because both have the same Lie algebra. This symmetry is global
and it does not necessary to introduce gauge fields.

When the symmetry is broken, the scalar field $\phi _{4}$ get a v.e.v
different from zero, and it can be redefined as follows 
\begin{equation}
\phi _{4}=H+v
\end{equation}
where $H$ gets its mass and is called the Higgs particle. Moreover, it has a
v.e.v equal to zero. This field is a physical degree of freedom and its mass
is proportional to the $\lambda $ parameter which is unknown in the model.
The other scalar fields remain massless. They are the would-be Goldstone
bosons and correspond to the degrees that the gauge fields `eat` in order to
get mass or longitudinal component.

The Higgs potential can be written as a function of the new fields as
follows 
\begin{equation}
V=\frac{\lambda }{4}\left( \phi _{1}^{2}+\phi _{2}^{2}+\phi
_{3}^{2}+H^{2}+2Hv+v^{2}-2\mu ^{2}\right) ^{2}\;.
\end{equation}
In this new potential the global symmetry is broken to $SO(3)$, which only
rotates three scalar fields. It is isomorphic to $SU(2)_{V}$, the diagonal
part of $SU(2)_{L}\otimes SU(2)_{R}$. We can say that the symmetry was
broken according to the following scheme 
\begin{equation}
SU(2)_{L}\otimes SU(2)_{R}\longrightarrow SU(2)_{V}.
\end{equation}
To evidentiate the isomorphism between $SO(4)$ and $SU(2)_{L}\otimes
SU(2)_{R}$, we will express the fields according to 
\begin{eqnarray}
\phi ^{^{\prime }} &=&U_{L}\;\;\phi \;\;U_{R}^{\dagger }\;,  \nonumber \\
\phi &=&\frac{1}{\sqrt{2}}\;\;\left[ 
\begin{array}{rr}
\phi _{4}-i\phi _{3}\;\; & \;\;\phi _{1}+i\phi _{2} \\ 
-(\phi _{1}-i\phi _{2})\;\; & \;\;\phi _{4}+i\phi _{3}
\end{array}
\right] \;,
\end{eqnarray}
where $U_{L(R)}\in SU(2)_{L(R)}$. The elements of the $SU(2)_{L}$ group act
over the columns and the elements of the $SU(2)_{R}$ group act over the
rows. These transformations leave invariant the following quadratic form 
\begin{equation}
Tr(\phi ^{\dagger }\phi )\;,
\end{equation}
and, obviously, leave invariant the Higgs potential.

If $\theta_L \! = \! \theta_R$ the group of transformations becomes $SU(2)_V$%
, and if $\theta_L\! =\! -\theta_R$ the group of transformations reduces to $%
SU(2)_A$. Under the $SU(2)_V$ infinitesimal transformations the fields of
the multiplet transform as follows 
\begin{eqnarray}
\phi^{^{\prime}}&=&e^{i \vec{\theta} \cdot \vec{T}} \; \phi\; e^{-i \vec{%
\theta}\cdot \vec{T}} \; ,  \nonumber \\
\delta\phi_i &\simeq& \epsilon_{ijk} \theta_j T_k \;\; , \; \; i, j, k=1, 2,
3 \; , \\
\delta\phi_4 &=& \delta H= 0 \; .  \nonumber
\end{eqnarray}

After the symmetry breaking, the quadratic form still invariant under $%
SU(2)_{V}$ is 
\begin{equation}
\phi _{1}^{2}+\phi _{2}^{2}+\phi _{3}^{2}=\phi ^{+}\phi ^{-}+\phi ^{-}\phi
^{+}+\phi _{3}^{2}\;,
\end{equation}

The interactions between fermions and the Higgs fields are given by the
Yukawa Lagrangian. It also gives mass to the fermions when the vacuum is
aligned. The renormalizable Yukawa Lagrangian is giving by 
\begin{equation}
L_{Y}=h_{u}\;Q_{L}\;\tilde{\phi}\;u_{R}+h_{d}\;Q_{L}\;\phi \;d_{R}+h.c.\;,
\end{equation}
where $Q_{L}^{T}=(u_{L}\;d_{L})$, $m_{u}=h_{u}<\phi _{o}>$ and $%
m_{d}=h_{d}<\phi _{o}>$. If we assume that the quarks have the same masses, $%
h_{u}\!=\!h_{d}$, then the Yukawa Lagrangian can be written as 
\[
h( 
\begin{array}{rr}
\bar{u}_{L}\; & \;\bar{d}_{L}
\end{array}
)\left[ 
\begin{array}{rr}
\phi _{4}-i\phi _{3}\;\; & \;\;\phi _{1}+i\phi _{2} \\ 
-(\phi _{1}-i\phi _{2})\;\; & \;\;\phi _{4}+i\phi _{3}
\end{array}
\right] \left( 
\begin{array}{r}
u_{R} \\ 
d_{R}
\end{array}
\right) +h.c.. 
\]
To get $L_{Y}$ invariant under the global $SU(2)_{L}\otimes SU(2)_{R}$
symmetry, the quarks have to transform according to 
\begin{equation}
Q_{L(R)}^{^{\prime }}\;=\;U_{L(R)}Q_{L(R)}.
\end{equation}
After SSB the quarks get mass and the global symmetry is broken down to $%
SU(2)_{V}$. The mass terms are given by 
\begin{equation}
L_{m}=h(\bar{u}_{R}\;u_{L}+\bar{d}_{R}\;d_{L})+h.c.\;,
\end{equation}
which are invariant under $U_{L}=U_{R}=SU(2)_{V}$, i.e., we have the isospin
symmetry. It is the same custodial symmetry.

\section{Custodial symmetry and radiative corrections}

Considering the contribution of the first doublet of fermions $(u,d)$ to the
radiative corrections of the $\rho $ parameter, we have 
\begin{eqnarray}
\Delta \rho &=&\Pi _{ZZ}(0)-\Pi _{WW}(0),  \nonumber \\
\Pi _{ZZ}(0) &=&-\frac{3G_{F}}{8\sqrt{2}\pi ^{2}}(2m_{u}^{2}\log
m_{u}^{2}-2m_{d}^{2}\log m_{d}^{2}), \\
\Pi _{WW}(0) &=&-\frac{3G_{F}}{8\sqrt{2}\pi ^{2}}\left( m_{u}^{2}+m_{d}^{2}+%
\frac{2m_{u}^{4}\log m_{u}^{2}-2m_{d}^{4}\log m_{d}^{2}}{m_{u}^{2}-m_{d}^{2}}%
\right)  \nonumber
\end{eqnarray}
where $\Pi _{ii}(0)=\sum_{ii}(0)/M_{i}^{2}$ and $\sum_{ii}(0)$ is the
diagonal contribution to the self energy. Summarizing all terms yielded \cite
{seis} we get 
\begin{eqnarray}
\Delta \rho &=&\frac{3G_{F}}{8\sqrt{2}\pi ^{2}}F(m_{u},m_{d}),  \nonumber \\
F(m_{u},m_{d}) &=&m_{u}^{2}+m_{d}^{2}-\frac{2m_{u}^{2}m_{d}^{2}}{%
m_{u}^{2}-m_{d}^{2}}\log \frac{m_{u}^{2}}{m_{d}^{2}}.
\end{eqnarray}
In the limit $m_{u}\!=\!m_{d}$, i.$\!$ e., when the isospin symmetry or
custodial symmetry is valid the function $F(m_{u},m_{d})$ is equal to zero.
The radiative corrections to $\rho $ are different from zero when the
custodial symmetry is broken. For example, the third family of the SM where
the up and down quarks are replaced by the top and the bottom quarks,
respectively, the global symmetry is broken because $m_{t}\gg m_{b}$ and the
correction to $\rho $ is proportional to $m_{t}^{2}$ 
\begin{equation}
\Delta \rho =\frac{3G_{F}\;m_{t}^{2}}{8\sqrt{2}\pi ^{2}}.
\end{equation}

\section{Conclusions}

When the SM symmetry is broken by Higgs doublets, there
is a global symmetry that protects the mass relation of the $W$ and $Z$
fields, which transforms as the components of a triplet. This relation is
violated when this symmetry is not exact. If the isospin symmetry or
custodial symmetry is broken, then $\rho $ parameter may get radiative
corrections different from zero. To look for new physics beyond the SM it is
important to know if this symmetry is still exact.

We thank to COLCIENCIAS for financial support.

\end{document}